\documentclass[9pt,twocolumn,oneside]{osajnl}

\journal{pr} 

\setboolean{shortarticle}{false}

\title{Near-Infrared 3D Imaging with Upconversion Detection}

\author[1,2]{He Zhang}
\author[1,2]{Santosh Kumar}
\author[1,2]{Yong Meng Sua}
\author[1,2]{Shenyu Zhu}
\author[1,2,*]{Yu-Ping Huang}

\affil[1]{Department of Physics, Stevens Institute of Technology, Hoboken, NJ, 07030, USA}
\affil[2]{Center for Quantum Science and Engineering, Stevens Institute of Technology, Hoboken, NJ, 07030, USA}
\affil[*]{Corresponding author: yuping.huang@stevens.edu}

\begin{abstract}
We demonstrate a photon-sensitive, three-dimensional camera by active near-infrared illumination and fast time-of-flight gating. It uses pico-second pump pulses to selectively up-convert the backscattered photons according to their spatiotemporal modes via sum-frequency generation in a $\chi^2$ nonlinear crystal, which are then detected by electron-multiplying CCD with photon sensitive detection. As such, it achieves sub-millimeter depth resolution, exceptional noise suppression, and high detection sensitivity. Our results show that it can accurately reconstruct the surface profiles of occluded targets placed behind highly scattering and lossy obscurants of 14 optical depth (round trip), using only milliwatt illumination power. This technique may find applications in biomedical imaging, environmental monitoring, and wide-field light detection and ranging. 

\end{abstract}

\setboolean{displaycopyright}{true}

\begin{document}

\maketitle

\section{Introduction}
Three-dimensional (3D) imaging has been a long and actively pursued technology due to its important applications in medical diagnosis \cite{Kim:10, Satat2016, Shi2019}, remote sensing \cite{Liou:06, McCarthy:13}, facial recognition \cite{Kirmani14,Rehain2020, app9071458}, environmental monitoring \cite{su11010162, Ren:19}, and so on.
A handful of techniques and realizations have thus far been demonstrated, including those based on structured-light imaging \cite{app9071458, Ko:18}, light detection and ranging (LiDAR) with raster scanning \cite{Imaki:07,Kirmani14,Tseng:18, Rehain2020}, and stereophotogrammetry \cite{Heike2010}. Recently, the time-correlated single-photon counting technology has been deployed to boost the detection sensitivity \cite{Massa:98, Tobin:2021,Tachella:19}. In general, those active-illumination systems can generate 3D profiles of target object with higher accuracy than those based on passive sensing.




Meanwhile, infrared (IR) imaging and detection has been studied extensively in the past decade, which allows sensitive detection of many biomolecular and chemical signals \cite{Amrania:12, Nallala:14} compared to that of visible light \cite{Dam2012, PhysRevApplied.11.044013}. However, existing IR detection techniques are mostly based on thermal sensors, which have low sensitivity and high noise, even with cryogenic cooling \cite{ROGALSKI2002187}. On the other hand, visible detectors have much lower noise and high sensitivity without the need for cryogenic cooling. Instead of the direct IR imaging, the parametric frequency upconversion imaging \cite{Faris:94,Barh:19, Zhou2013} plays a critical role for hyper-spectral IR imaging, where the IR signal is frequency upconverted into visible wavelength \cite{Dam2012, Demur:18, Junaid:19, Single-Photon_Edge_Enhanced2021, Huang:21} and detected by a silicon-based detector or camera with high sensitivity and low noise. Many unique nonlinear optical systems have also been developed for 2D imaging, such as noise-less optical parametric amplification imaging \cite{PRL_Kumar, Frascella:21}, non-degenerate two-photon absorption \cite{Knez2020, Liu:21, Potma:21}, spontaneous parametric down conversion imaging \cite{Paterova_SA2020, Basset_2019}. Some are deployed to NIR or MIR imaging regimes. With parametric frequency up-conversion, it has been shown that near-unity conversion efficiency can be achieved, and also can preserve the quantum characteristics of infrared photons \cite{PhysRevLett.68.2153}. This will facilitate NIR or MIR imaging at a few photon levels with low dark noise \cite{Dam2012}. For example, by utilizing a sensitive detector such as silicon electron-multiplying charge-coupled device (EMCCD) to directly record upconversion photons in the visible or near-infrared region \cite{Barh:19, Zhou2013}.

The natural extension of the existing 2D parametric up-conversion imaging system to 3D imaging can have great potential, and may offer promising applications that require infrared multidimensional imaging. One attempt on 3D IR imaging with parametric upconversion process is shown in \cite{Tanaka:18}. Here they used a chirped ultrashort pulse as the laser source and utilized the principle of ultrafast conversion between space, time, and frequency to obtain a single shot classical 3D image. However, the weakness of their study was due to the trade-off between the measurement range and depth resolution, as well as the low spatial resolution due to tight focusing at the crystal plane. This makes it difficult to extract critical features of the objects. 

\begin{figure*}[htbp]
    \centering
    \includegraphics[width=\linewidth]{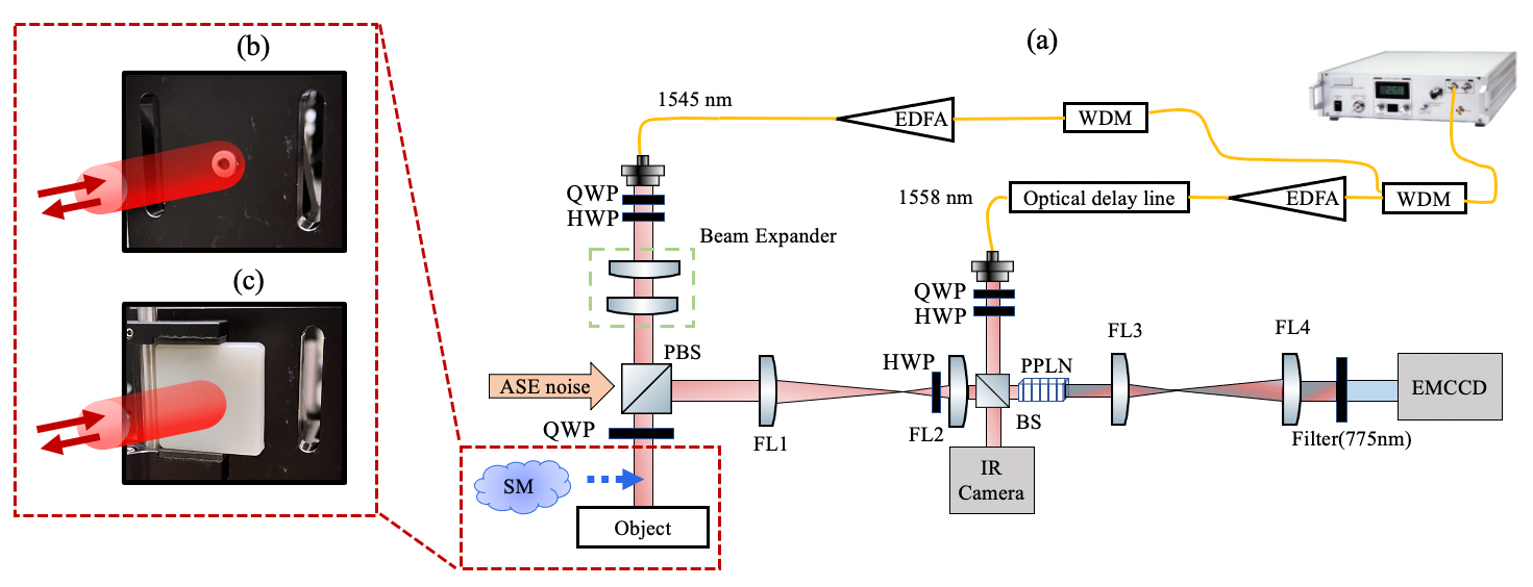}
    \caption{(a) Experiment setup. Mode-locked laser pulses are separated into two arms by using WDM filters, with signal and pump wavelengths at 1545 and 1558 $nm$, respectively. The signal beam is incident on an obscured object. The back-scattered signal photons are combined with the pump, then up-converted in a nonlinear crystal to generate SF output centered at wavelength 775.5 $nm$. The time-resolved measurements can faithfully reconstruct the 3D object image captured by the EMCCD camera. (b) The picture of the object (a washer) attached on an aluminum block. (c) The picture of the obscured object, i.e. the washer obscured by the scattering media (SM). WDM: wavelength division multiplexers, EDFA: Erbium-Doped Fiber Amplifier, QWP: Quarter Waveplate, HWP: Half Waveplate, BS: Beamsplitter, FL: Fourier lens, PPLN crystal: Magnesium-doped Periodic Poled Lithium Niobate crystal, EMCCD: electron multiplying silicon charged coupled device, ASE: amplified spontaneous emission. 
}
    \label{fig:setup}
\end{figure*}

Some common challenges in most of the back-reflected imaging systems are strong background noise and multiple scattering \cite{Dunsby_2003, Liu:21, Wu:17, Shi:19}. In order to strip the background noise from the contaminated signal, traditional methods such as time-frequency filters are commonly used \cite{940617,Shahverdi:18}, but inherently limited by the trade-off between signal detection efficiency and noise rejection. Realizing this, quantum parametric mode sorting has been proposed and demonstrated \cite{Shahverdi:18, Shahverdi-lidar, Zhang2019}. Some other optical techniques are also present in the literature to address imaging through multi-scattered media. Most of these techniques can be broadly classified as relying on ballistic photons \cite{Dunsby_2003, Cho:21}, diffuse optical tomography \cite{PhysRevX.11.031010, Radford:20}, etc. Those techniques offer potential applications for medical imaging, communications and security \cite{Satat2016, Katz2014}. However, most studies in the field of overcoming strong background noise and multiple scattering have focused on 2D imaging. In the past, QPMS has been utilized for single-pixel 3D single photon imaging system in photon-starved, noise-polluted environments as well as imaging through strongly scattering medium \cite{Rehain2020, Maruca:21}. With raster scanning methods, it demonstrated imaging of a 3D object through a highly reflective obscuring scene with 36 dB advantage in noise rejection. However, the requirement of raster scanning has severely impeded the image acquisition speed while the traverse spatial resolution of the 3D image is limited by the field of view of single pixel detector \cite{feihu20}.






In this paper, we extend these promising studies and explore an active NIR 3D imaging system using EMCCD with upconversion detection through a highly scattered medium amid strong spatio-temporal background noise. Our system is based on the nonlinear frequency upconversion process via sum-frequency (SF) generation of time-correlated pump and signal optical pulses. Combined with single-photon sensitive EMCCD, we capture spatial and temporal information of the scene of interest, which can be used to reconstruct a 3D surface profiles of the target object. To surmount the imaging through strong scattering medium, we utilize time-resolving photon detection by detecting the frequency up-conversion image confined temporally within the time window of the pump pulse, which shows excellent behavior in noise rejection. Compare with the raster scanning methods \cite{Rehain2020, Imaki:07, Tseng:18}, in our work, we could significantly improve the spatial resolution up to 48 $\mu m$ and effectively reduce the 3D image acquisition time. Besides, we could reconstruct the 3D profile of the target object through strong spatio-temporal background noise (SNR about -20 dB). This system can be deployed in applications that require ultra-sensitive imaging, such as medical diagnostics and quantum optics at single or few-photons level \cite{Brida2010, PhysRevLett.110.153603}, and find values in biomedical imaging, non-destructive label free diagnosis and quantum communications. In the future, selective 3D edge enhancements can be implemented by imprinting the spatial phase patterns on the pump beam using a spatial light modulator (SLM) \cite{Qiu:18, Li:20}.
 


.  




\section{Experiment setup}
\label{sec:setup}

\begin{figure*}[htbp]
    \centering
    \includegraphics[width=0.85\linewidth]{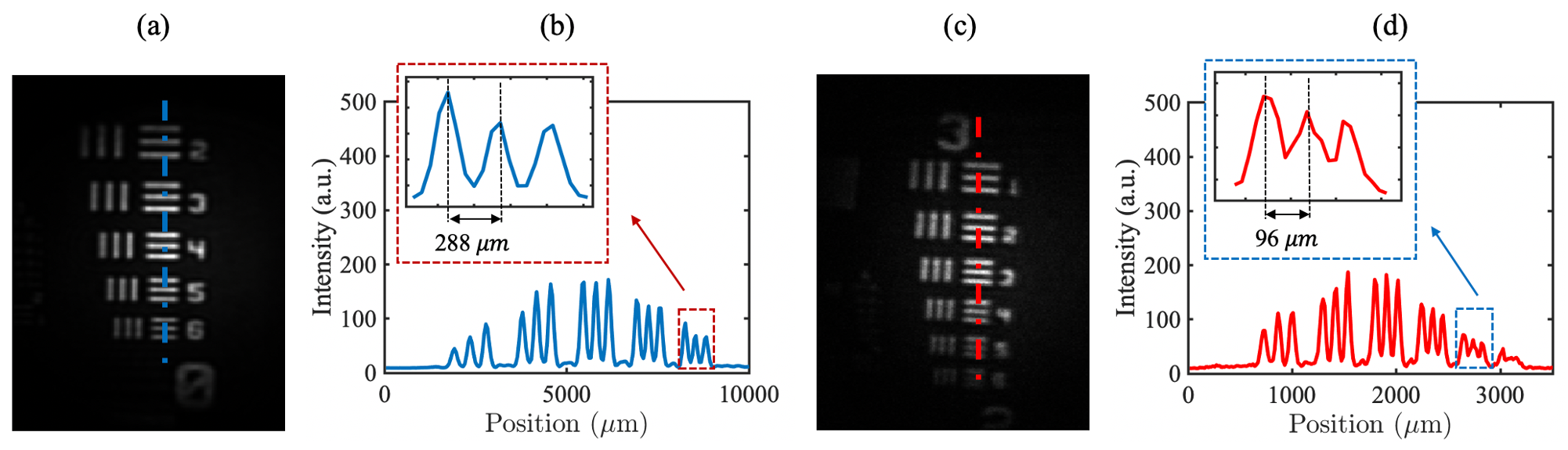}
    \caption{Field of view and spatial resolution of upconverted images using two different experimental situations. (a) The up-conversion image of group 1 in USAF resolution test chart from the experiment setup depicted in Fig. \ref{fig:setup}(a). (b) The intensity profile along the blue dashed line in (a).  (c) The up-conversion image of group 3 in USAF resolution test chart with another experimental settings. In this case, the signal beam size is reduced to 3.6 $mm$ FWHM and the first Fourier lens(FL) of the object 4f imaging system is changed to F1 = 100 $mm$. (d) The intensity profile along the red dashed line in (c). }
    \label{fig:spatial resolution}
\end{figure*}


The nonlinear optical setup for the 3D image reconstructor is shown in Fig. \ref{fig:setup}(a). The signal and pump pulses are derived from a 50 MHz femtosecond mode-locked laser (MLL) using two inline narrowband wavelength division multiplexers (WDM) of 0.8 $nm$ linewidth to pick two separate wavelengths, one at 1545 $nm$ as the signal and another at 1558 $nm$ as the pump. The pump pulses are sent to a programmable optical delay line (ODL) and then guided into free space with collimated beam size of 0.65 $mm$ FWHM. The collimated signal beam is magnified by a beam expander to 10.8 $mm$ FWHM. The intensity of the horizontally polarized signal beam is tuned by the combination of a half and a quarter waveplates (HWP $\&$ QWP) with a polarizing beam splitter (PBS). The signal light, after PBS, then passes through another QWP and scattering media before incident on the object. The back reflected or scattered light from the object then changes the polarization 
and pass through a telescope with lenses FL1= 300 $mm$ and FL2 = 25 $mm$ which reduce the beam size to $\sim$ 0.53 $mm$ FWHM. After that, a beam splitter (BS) combines the collimated signal and pump beams which incident into a temperature-stabilized PPLN crystal with the poling period of 19.36 $\mu$m (5 mol.\% MgO doped PPLN, 2 $mm$ length, 3 $mm$ width, and 1 $mm$ height from HC Photonics) for frequency conversion process. The normalized conversion efficiency in our case is $9\times 10^{-4}\%/W$, which is restricted by three factors: (a) low pump power, (b) short crystal length (2 $mm$) and (c) not in the optimum focusing condition for signal and pump (both are collimated beam) inside the crystal. 
The 4f system after the crystal consists of two Fourier lenses with focal length FL3 = 25 $mm$ and FL4 = 100 $mm$, imaging the SF output onto a EMCCD (iXon Ultra 897, Andor) with $512 \times 512$ pixels and 16 $\mu$m pixel size. The quantum efficiency of this EMCCD is measured to be $7.5\%$ at upconversion wavelength, calibrated against a silicon avalanche photo diode at mean photon number $\approx$ 0.01 per pulse. On the other output of the BS, an IR camera (FIND-R-SCOPE Model No. 85700) is used to capture the image via direct signal detection.






\section{results}

First, we evaluate the transverse spatial resolution of the 3D imager with two different experimental settings, as shown in Figure \ref{fig:spatial resolution}. A 1951 USAF resolution test chart (USAF-RTC) is used at the object position to test the spatial resolution of this system. There are 54 target elements provided in the USAF-RTC, and each element consists of three bars which are separated by the bar width. We define the feature size as the width of the bar in the USAF-RTC, which is half the distance between the centers of the two bars. The spatial resolution result shown in Fig. \ref{fig:spatial resolution}(a) is the up-conversion image of group 1 in USAF-RTC, which was obtained from the experimental setup depicted in Fig. \ref{fig:setup}(a). The intensity vs position line plot shown in Fig. \ref{fig:spatial resolution}(b) reflects the 1D spatially resolved blue dashed line in Fig. \ref{fig:spatial resolution}(a). It shows that our setup is able to easily resolve features with size $\sim$144 $\mu m$, which is also consistent with its actual value 140.31 $\mu m$ (group 1, element 6). The decrease in the intensity at the edges is due to the Gaussian intensity  distribution of the probing signal beam and the pump beam. We have validated that the size of the spot hitting on the object and the subsequent imaging system affect the final resolvable resolution of our 3D imager system, since they affect the point spread function of back reflected signal from the object. So, another version of the experiment setup has been constructed, whose spatial resolution result is shown in Fig. \ref{fig:spatial resolution}(c), which is the up-conversion image of group 3 in USAF-RTC. In this case, the signal beam size is reduced to 3.6 $mm$ FWHM. Besides, the FLs of the first 4f system are replaced as FL1 = 100 $mm$ and FL2 = 25 $mm$. The intensity profile along the red dashed line in Fig. \ref{fig:spatial resolution}(c) is shown in Fig. \ref{fig:spatial resolution}(d), where we can easily resolve objects with spatial resolution 48 $\mu m$. This feature size is matched with its actual value 49.37 $\mu m$ (group 3, element 5). In this case, the resolvable feature size is improved, but the field of view from the observed object is reduced. 
In order to image bigger objects, we present the following results in this work using the setup shown in Fig. \ref{fig:setup}(a), with the spatial resolution of 144 $\mu m$. 

\begin{figure*}
    \centering
    \includegraphics[width=\linewidth]{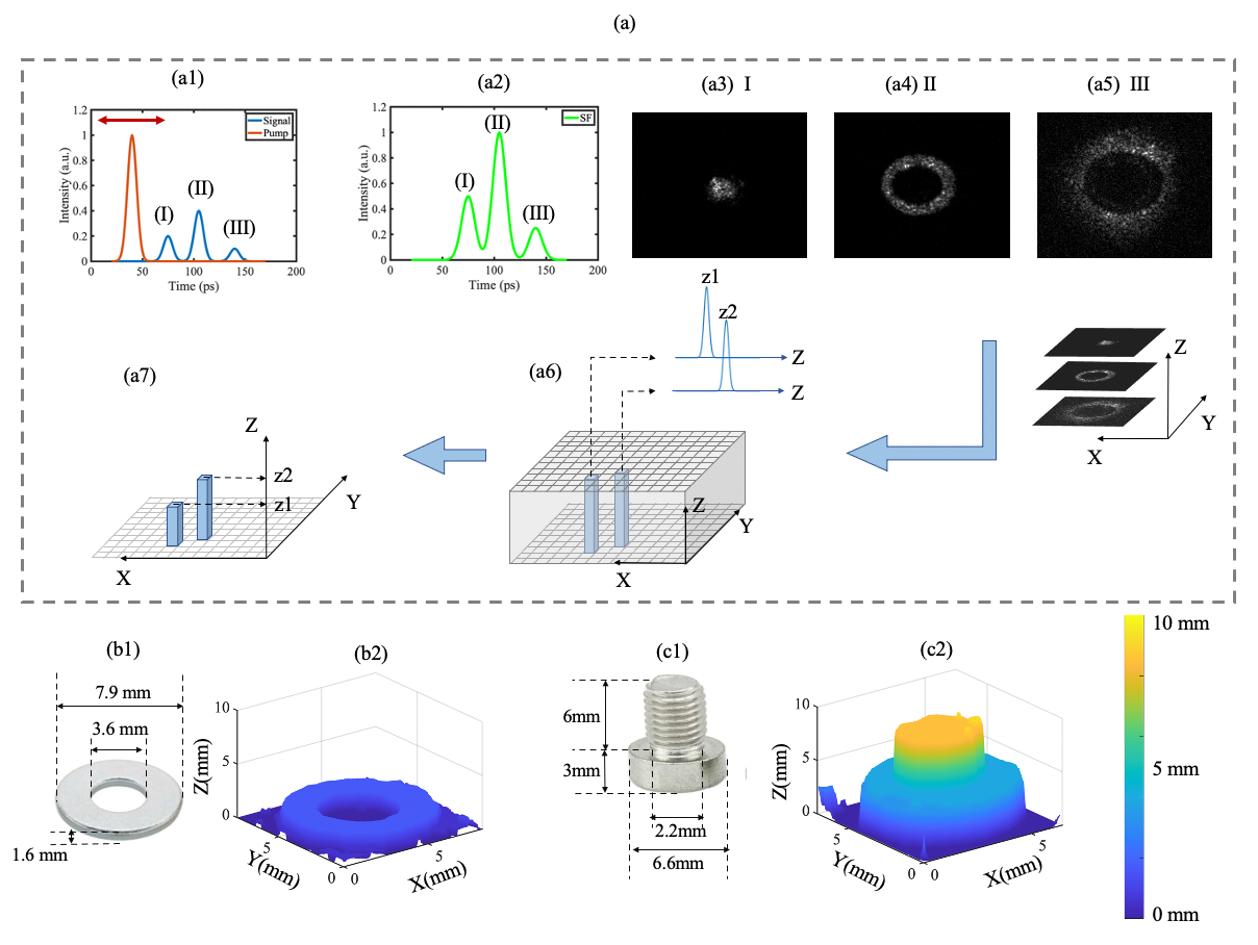}
    \caption{(a1)-(a7) illustrate the present 3D imaging method. (a1) presents the intensity measurement for the input signal and pump at different arrival times, with the corresponding SF intensity shown in (a2). (a3)-(a5) shows the spatial information at different arrival time $\uppercase\expandafter{\romannumeral1}$, $\uppercase\expandafter{\romannumeral2}$, and $\uppercase\expandafter{\romannumeral3}$, respectively. (a6) shows the reconstruction procedure of the 3D data set collection. At each pump delay, 2D image data are acquired by the camera in each frame, and the $z$-dimension represents the photon's fight distance. (a7) shows post-processing data of (a6), in which the $z$-axis gives the depth information of the target object. The object photos with profile data for washer and bolt are shown in (b1) and (c1), respectively. After performing 3D imaging measurement, the results for washer and bolt are shown in (b2) and (c2), respectively. } 
    \label{fig:no-scatter}
\end{figure*}

Our 3D object image reconstruction technique utilizes the spatial and temporal photon information enabled by the nonlinear frequency conversion process. Figure \ref{fig:no-scatter} shows an example of the 3D reconstruction process. In Fig. \ref{fig:no-scatter} (a1), the blue curve indicates the normalized photon intensity as a function of the arrival time for the reflected/back-scattered signal photons from the object to reach the crystal, and the red solid curve represents the pump pulse arrival time. By sweeping the optical delay line, the pump pulses are swept along the temporal domain, which can overlap with the returning signal pulses at a certain arrival time. These two overlapped pulses interact inside the PPLN crystal, and the SF light is generated as shown in the Fig. \ref{fig:no-scatter} (a2). The signal can be up-converted efficiently only if it is spatially and temporally overlap with the pump. At each delay step, the SF image was captured to reconstruct the depth $z$ ($=c\times t/2$) of the 3D object. Therefore, the optical delay time indicates the arrival time of the back-scattered signal photons. In Fig. \ref{fig:no-scatter} (a3)-(a5), three SF images collected at optical delay time t = 75 $ps$, t = 105 $ps$, and t = 140 $ps$ respectively. 
The collected data is a 3D data set shown in Fig. \ref{fig:no-scatter} (a6). Each pixel has one histogram of photon's arrival time versus the counts. After converting photon's arrival time into distance, we show two histograms for two different pixels in Fig. \ref{fig:no-scatter} (a6). The peak value (z1 and z2) of the curve provides the information of relative depth for each pixel on the object. By post-processing the data, the reconstructed image with depth information is shown in Fig. \ref{fig:no-scatter} (a7) followed by applying a median filter (4 $\times$ 4 pixels) to smoothen the reconstructed image. 
In the experiment, a washer and a bolt are used as the target objects to perform the measurement. The dimensional measurement of the washer and the bolt are shown in Fig.\ref{fig:no-scatter} (b1) and (c1), respectively. The results of the reconstructed 3D images are shown in Fig.\ref{fig:no-scatter} (b2) and (c2). The X-axis and Y-axis give the cross-section of the target object in $mm$, and Z-axis shows the depth information of the target object in $mm$. In the Fig.\ref{fig:no-scatter} (b2), the measured outer diameter of the reconstructed washer is 7.92 $mm$, inner diameter is 3.65 $mm$, and depth is 1.6 $mm$. In the Fig.\ref{fig:no-scatter} (c2), the measured stub height and diameter of reconstructed bolt is 5.9 $mm$ and 2.3 $mm$,  the height and the diameter of the bottom parts is 3.2 $mm$ and 6.45 $mm$. We set the EMCCD exposure time of each image to 1 s. The total acquisition time for a full 3D image reconstruction is $\sim$30 s. The reconstructed 3D image in Fig.\ref{fig:no-scatter} (b2) and (c2) agree well with the ground truth shown in (b1) and (c1). 



\begin{table}[htbp]
\centering
\caption{The parameters of scattering media.}
\begin{tabular}{cccc}
\hline
  & Sample  & Mean free path & Optical depth  \\
    &  &($l_s$) &  (round trip) \\
\hline
SM1 & 4.3 mm & 0.58 mm & 14.58$l_s$ \\
SM2 & 3.2 mm & 0.66 mm & 8.08$l_s$ \\
\hline
\end{tabular}
  \label{tab:scattering sample}
\end{table}



\begin{figure*}[htbp]
    \centering
    \includegraphics[width=0.9\linewidth]{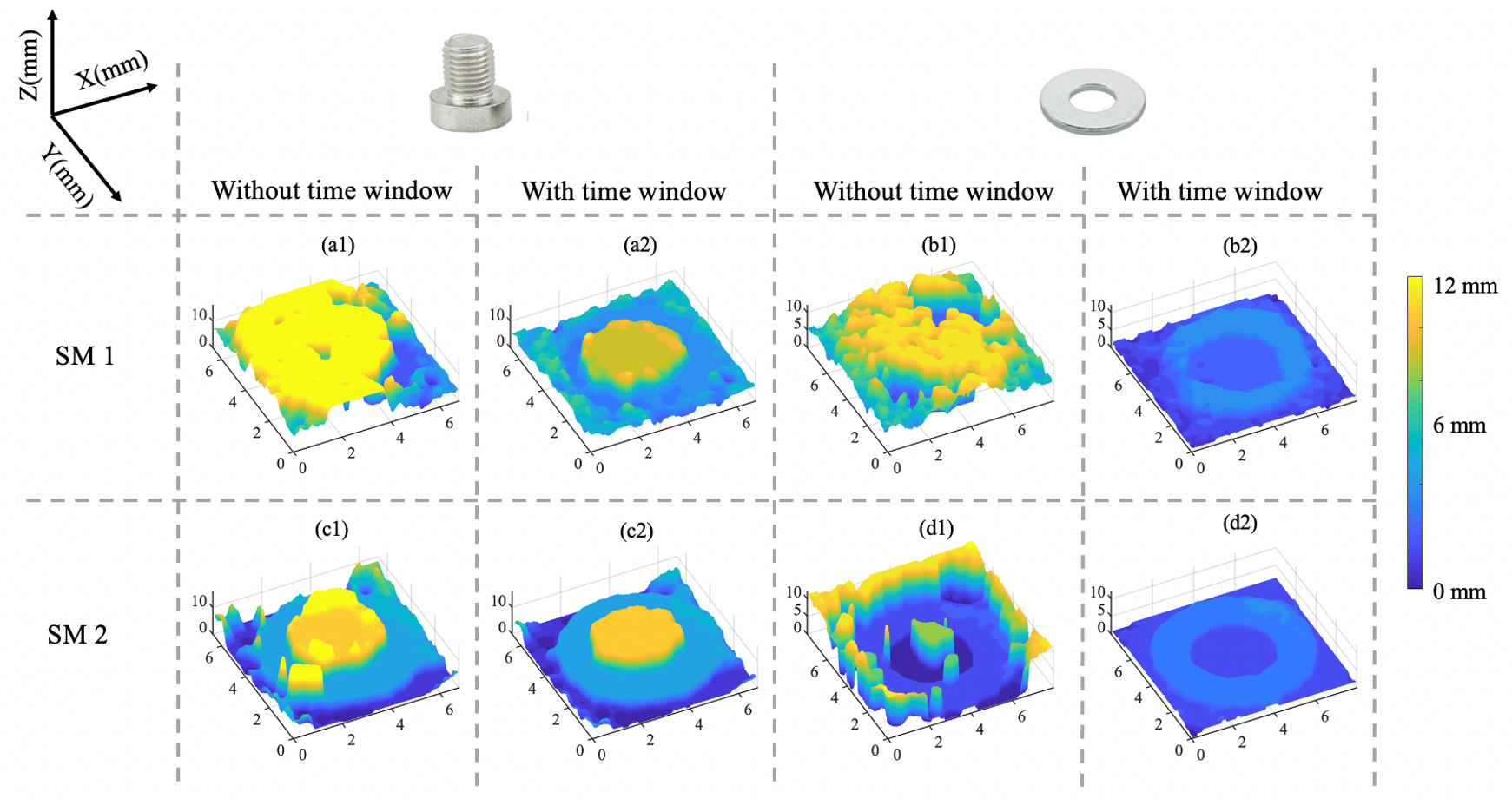}
    \caption{3D reconstructed image through scattering media. Two SM samples are used (Table \ref{tab:scattering sample}), the results of SM1 with optical depth 14.58$l_s$ for double passes shown in second row, and the results of SM2 with optical depth 8.08$l_s$ shown in third row. The time window can partially discard the redundant noise in the temporal scans and improve the reconstructed image contrast.   }
    \label{fig:scatter}
\end{figure*}

Next, we test the performance of our technique through scattering media. The scattering media are made from epoxy resin and Titanium Oxide (TiO2) pigment ($\sim $220 nm particle size). We examine two different pieces of scattering media (SM1 and SM2) in our setup. The thickness, mean free path ($l_s$) and optical depth of the scattered media are shown in Table \ref{tab:scattering sample}. SM1 with thickness 4.3 mm and optical depth 7.29$l_s$ (14.58$l_s$ round trip) is more scattered compare to SM2 with thickness 3.2 mm and optical depth 4.04$l_s$ (8.08$l_s$ round trip). 
In our case, the scattering media is placed in front of the object, so the signal propagates twice through the scattering media before upconversion detection. 
The back-reflected photons coming from the different surface of the target object are upconverted in different time intervals using the pump pulse time gating. It could allow us to isolate the other back-scattered noise photons arriving in different time intervals. 
Figure \ref{fig:scatter} shows the 3D image reconstruction of the target objects, bolt ((a) and (c))  and  washer ((b) and (d)) through scattering media (SM1 and SM2), respectively. To effectively reconstruct the 3D image, we used a time windowing procedure to post-select several continuous EMCCD images that are captured at different temporal delay of the pump pulse. The time window could discard most of the background noise coming from the scattering, and process only those within the time window defined by the pump pulses,  following the procedure shown in Fig.\ref{fig:no-scatter}(a6) to (a7). This procedure is programmed via Matlab to improve the SNR of the reconstructed 3D image. It reduces the speckle noise induce by the scattering medium, thus better resolve the shape of the target object. The reconstructed image of the bolt through SM1 without and with time windowing are shown in Fig.\ref{fig:scatter} (a1) and (a2), respectively.  
Here, the edge and depth variation of the cap bolt can be distinguished clearly. The washer image without and with temporal windowing are also reconstructed through SM1 in Fig. \ref{fig:scatter} (b1) and (b2), respectively. It can also reconstruct the image of the washer. 
Similar results are shown in the third row of Fig. \ref{fig:scatter} for relatively weaker scattering media (SM2). In both cases, we can effectively reconstruct the 3D image of the target object by carving the temporal window.

\begin{figure*}[htbp]
    \centering
    \includegraphics[width=0.8\linewidth]{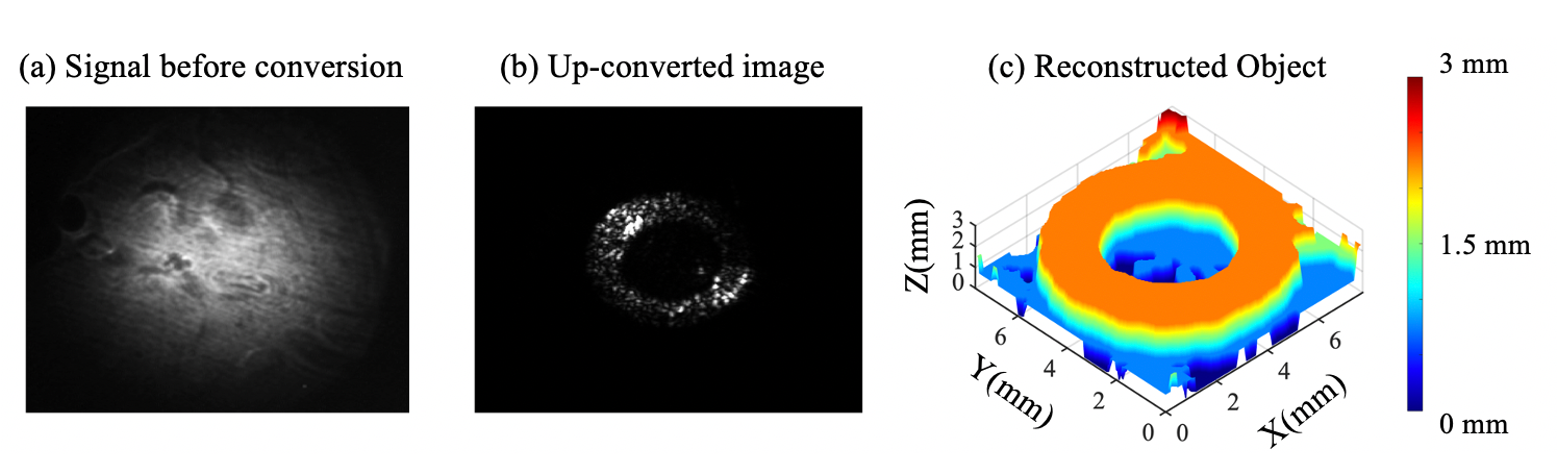}
    \caption{3D image reconstruction through addition noises in the time-frequency and spatial domain. (a) The signal image before up-conversion captured using the IR camera, which reflect from the target object and mixes with time-frequency and spatial noises. (b) The up-converted SF image at a certain arrival time, which can captured using the EMCCD. (c) The reconstructed 3D image. }
    \label{fig:noise}
\end{figure*}

After that, we inject the amplified spontaneous emission (ASE) noise that is temporally and spectrally overlapping with the signal, as shown in \ref{fig:setup} (a). This noise is generated from an erbium doped fiber amplifier (EDFA). Both the signal and ASE noise pass twice through SM2 before the upconversion detection. To ensure that the ASE noise is in the same time-frequency and spatial profile, we choose the same WDM filter bandwidth and spatial beam size, as the signal. In this measurement, we use the washer in Fig.~\ref{fig:no-scatter} (b1) as the target object. Figure~\ref{fig:noise} (a) shows the image of the signal mixed with the temporal noise before up-conversion, as taken by the IR camera. At a low SNR (about -20 dB), it is impossible to reconstruct the image of the target object with direct detection. Yet, our system captures an image with up to $8\times 10^7$ converted photons per second using the EMCCD as shown in Fig.\ref{fig:noise} (b). By time windowing, the noise is effectively suppressed, and the reconstructed washer image gives $\sim$16 dB improvement in the SNR. Figure \ref{fig:noise}(c) shows the 3D reconstruction image of the target object by temporally scanning the pump pulses. The exposure time of the EMCCD is the same as the other previous measurements (= 1 s). This result clearly demonstrates the noise rejection advantage of our system. 

\begin{figure*}[htbp]
    \centering
    \includegraphics[width=0.8\linewidth]{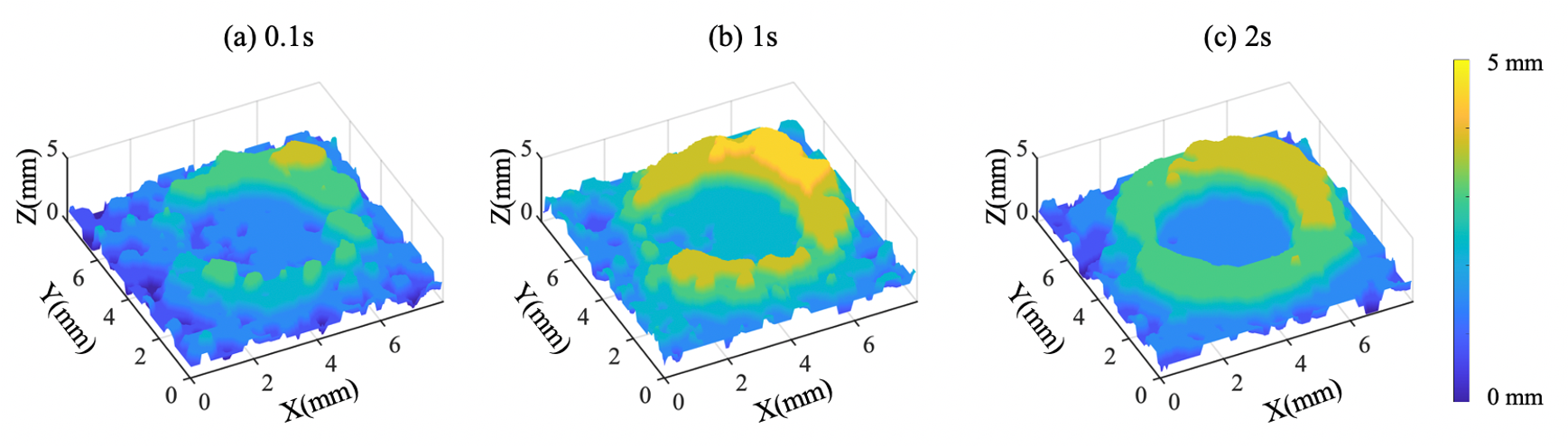}
    \caption{Reconstructed 3D image of washer for a series of exposure times (a) 0.1 s, (b) 1 s, and (c) 2 s.}
    \label{fig:expose time}
\end{figure*}
We now briefly discuss the effects of the EMCCD exposure time on the reconstructed 3D images. Figure~\ref{fig:expose time} shows the results for three different exposure times: 0.1 s, 1 s and 2 s. As the exposure time increases, the 3D image of the washer can be easily recognizable with better contrast. For the exposure time of 2 s on EMCCD, the total acquisition time in reconstructing a 3D image is $\sim$ 60 s, which is considerably long. Yet, as the above improvement comes from the increased number of photons collected by the EMCCD, one can instead increase the nonlinear conversion efficiency or use detector arrays with higher quantum efficiency, to enhance the image contrast while reducing the acquisition time.


\begin{figure*}[htbp]
    \centering \includegraphics[width=0.7\linewidth]{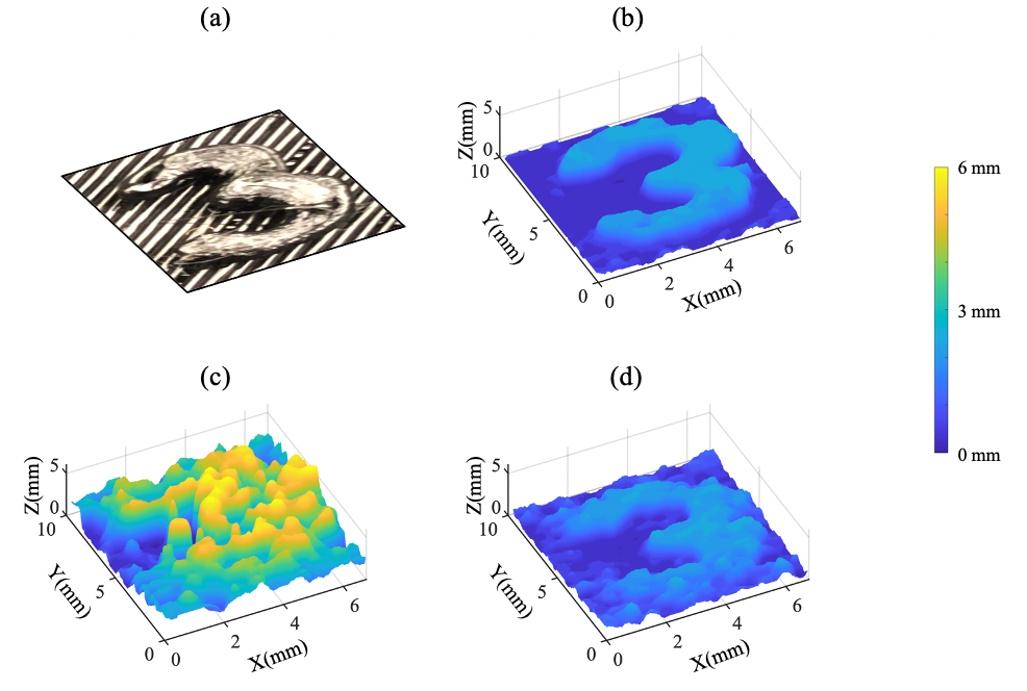}
    \caption{3D imaging measurement for the target object with diffusive surface. (a) A photo for target object. (b) A reconstructed image without scattering media. (c) A reconstructed image of the object with scattering media placed in front of it. (d) Post processing image of (c).}
    \label{fig:3d pinter}
\end{figure*}

Thus far, we have used highly-reflective metal objects as our testing targets. To further assess the capability of our photon sensitive 3D imager for general objects, we now switch to a target with diffusive surface, shown in Fig.~\ref{fig:3d pinter}(a). This diffusive surface, a digit '3', is homemade using PLA plastic Filament by 3D printer. For 3D reconstruction of this object, we set the EMCCD exposure time to 2s. Figure~\ref{fig:3d pinter}(b) shows the reconstructed 3D image without any scattered media or ASE noise. It shows that the 3D reconstructed image is clearly recovered. When we add the same scattering media and ASE noise as used in Fig.~\ref{fig:noise}, our 3D imager performance degrades. Nonetheless, we are still able to recover the object by properly choosing the temporal window. Figure \ref{fig:3d pinter}(c) presents the results with the full time scanning, while Fig.~\ref{fig:3d pinter}(d) gives the manicured data by carving the temporal window. In these cases, we need to increase the EMCCD exposure time to 10 s to collect more photons. As seen, the unrecognizable 3D image in (c) can be well recovered in (d) by properly tuning the temporal window. 


\section{ Conclusion}
We have experimentally demonstrated a high-performance 3D imager for photon-sensitive detection using optical frequency upconversion pumped by picosecond pulses. 
It obtains millimeter depth resolution and 140 $\mu m$ spatial resolution, while effectively rejecting background noise from ambient environment and obscurants. As such, the present technique could potentially find applications in biomedical imaging, remote sensing over low visibility.
On the other hand, in the current method, the acquisition time of photon-sensitive imaging is longer than what is needed for typical real-time target object identification. This shortcoming is mainly due to the low conversion efficiency of the current nonlinear process, which can be increased by using a high-power laser or a longer nonlinear crystal. Also, to improve the imaging sensitivity through different kind of scattered materials, one could use spatially modulated pump beams to further improving the 3D image contrast, similar to demonstrated in 2D imaging cases \cite{Qiu:18, Li:20}. 


\,~~~~\,

\noindent\textbf{Acknowledgement.}
This material is based upon work supported by the ACC-New Jersey under Contract No. W15QKN-18-D-0040.

\,~~~~\,

\noindent\textbf{Disclosures.} The authors declare no conflicts of interest.

\,~~~~\,

\noindent\textbf{Data Availability.} Data underlying the results presented
in this paper are not publicly available at this time, but may
be obtained from the authors upon reasonable request.

\,~~~~\,

\bibliography{sample}



\end{document}